# Strong angular dependence of resonant states in 2D photonic molecules

Ángel Andueza,[1,*] Jesús Pérez-Conde,[2] Joaquín Sevilla[1]

[1]Dpto. Ing. Eléctrica y Electrónica Universidad Pública de Navarra, 31006 Pamplona, Spain. Smart Cities Institute, Universidad Pública de Navarra, 31006 Pamplona, Spain
[2] Dpto. de Física Universidad Pública de Navarra, 31006 Pamplona, Spain
*Corresponding author: angel.andueza@unavarra.es

**Photonic Molecules made of dielectric cylinders assembled as two-dimensional octagons and decagons have been studied. These structures exhibit resonance states that change their intensity depending on the incident radiation angle. While most part of spectra present small or even null variation, one of the resonances in both structures presents a high-sensibility to the incidence angle. This strong variation is well described in terms of the electric field intensity distribution of this resonant state, which is composed of maxima with alternate polarities inside each cylinder. Experimental measurements in the microwave range were taken from photonic molecules prototypes made of centimeter-scale glass cylinders ($\varepsilon$=4.5). The excellent agreement between measurements and simulations as the incidence angle varies makes this system ideal to develop angle sensors.**

*OCIS codes: (140.3490) Lasers, distributed-feedback; (060.2420) Fibers, polarization-maintaining; (060.3735) Fiber Bragg gratings; (060.2370) Fiber optics sensors.*

http://dx.doi.org/10.1364/OL.99.099999

Photonic crystals and, later, metamaterials marked a breakthrough in light control and manipulation. Photonic crystals (PhCs) are created with different structures: regular [1,2], quasicrystalline [3,4] or disordered [5,6] of the dielectric constant over distances of the same order of the light wavelength. Following the analogy with molecules made from atoms, smaller structures made of resonators or "photonic atoms" can be built resulting in Photonic Molecules (PM), an idea extensively explored [7-16]. Rings of cylinders with n-fold rotational symmetries appear in 2D Quasi-Periodical (QP) PhCs [16,17,18]. These rings, that can be considered photonic molecules, are responsible of discrete states inside complete photonic band gaps in QP PhCs structures [16]; these molecular states arise from the combination of isolated Mie resonances of the individual cylinders [4]. To our knowledge, an analysis of these high-symmetry local centers on their own, isolated from QP PhCs structures, has not been performed previously. This is the aim of this paper.

This work explores the electromagnetic properties and applications of PMs based on dielectric rods with octagonal and decagonal symmetries, paying special attention to the rotation angle of PMs with respect to the incident radiation direction. In order to complete the understanding of the basic behavior of these systems, we have numerically studied their scattering response and electric field distribution. Octagonal and decagonal clusters of glass cylinders were built (see Fig. 1), and their microwave transmission spectra were measured for different incidence angles. Two modes are clearly observed in the transmission response as the structure is rotated, one of them sensitive to angle rotation while the other is almost independent of it. These modes could be used as differential sensor where the angular position of the structure would be determined by the difference of the transmission intensity between them.

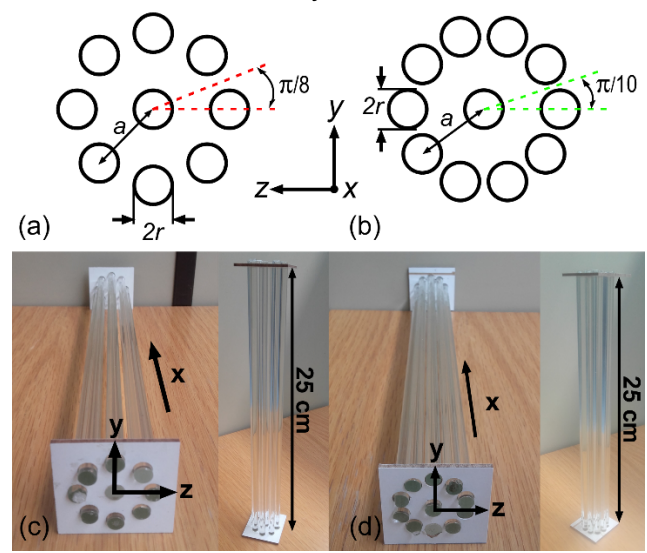

Fig. 1. Axis labeling and definition of the design parameters in (a) and (b). Photonic molecules studied: (c) octagonal and (d) decagonal clusters of cylinders built.

The selected structures are shown in Figure 1; the cylinders are placed in both configurations circularly around a central axial cylinder. The angular distance between cylinders centers is $2\pi/N$, where N is the number of cylinders of the ring and $a$ is the radial distance with the central cylinder. We have performed the measurements and calculations in the interval between the two extreme cases of azimuth angle: at 0° and 18° (22.5°), as the situation repeats passed these values due to rotational symmetry.

Numerical calculations were carried out with CST MICROWAVE STUDIO, a commercial code based on the Finite Integration time-domain Method (FIM) [19]. This program is an electromagnetic field simulation software package especially suited for analysis and design in the high-frequency range.

The prototypes to be measured in the microwave range (see Figure 1) were built using soda-lime glass cylinders of radius $r$=3 mm (with a dispersion of less than 3%) and length L= 250 mm. Cylinders were held parallel to each other and fixed between two wooden pieces where their positions were previously carved with a numerical control drill. The radial distance in both cases was $a$=12 mm and the aspect ratio of the structure, defined as AR=L/2$r$, was 33. This value assures a negligible influence of the finitude in the x-direction on the measurements, what is attained for AR>10 [20]. Transmission spectra of the samples were measured using a vector network analyzer (VNA) HP 8722ES spliced to rectangular horn antennas 79×46×129 mm (Narda Model 640) in a set up sketched in figure 2 [15,21].

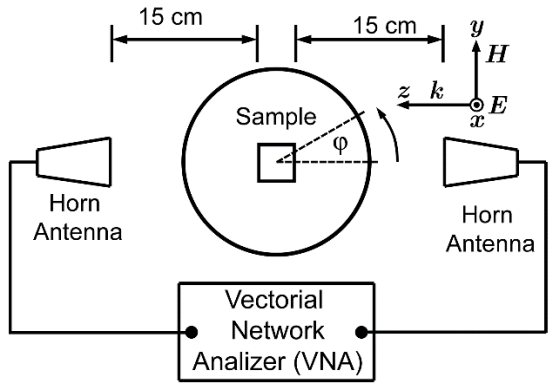

Fig. 2. Outline of the experimental setup. Radiation is emitted and detected in the perpendicular direction to the plane of the sample.

We first calculate the scattering cross section (SCS) of the two structures, the results are shown in Figure 3. Frequency is presented as the normalized parameter ($a/\lambda$) as well as in the actual frequency units (GHz). Two lines are plotted for each structure representing the SCS when the orientation of the structures with respect to the incident radiation is in the two extreme angular values (see Figure 1).

Both PM's present scattering spectra with very similar behavior. For low frequencies, the SCS of the PM is identical for the two extreme orientations. At higher frequencies, three peaks can be identified (labeled A, B, and C in the figures) whose value changes for the two angles represented. When rotating the structure, the first broad peak (A) remains almost inalterable, meanwhile, the second one (B) disappears totally. The third peak (C), is not present at 0° and grows for the increased angle.

In order to obtain a more detailed view of the resonance states of the structures, we computed the spatial distribution of electric field for peaks A and B. In the calculation, polarized radiation in the x-direction with unit amplitude propagating in the z-direction was used. The electric field distributions of the peaks A and B in plane yz for both, octagonal and decagonal PMs, are depicted in Fig. 4 and Fig. 5, respectively. In both cases, peak B undergoes a critical reduction of the electric field in comparison with the structure at 0°. The influence of rotation is weaker for peak A, where the magnitude of the electric field is similar in the two angles presented.

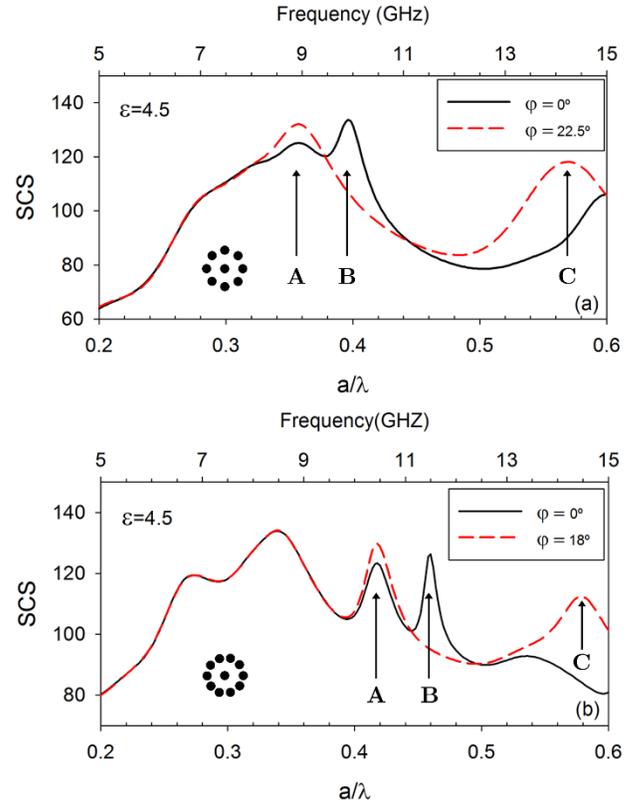

Fig. 3. Scattering Cross Section (SCS) as a function of the normalized frequency for the (a) octagonal and (b) decagonal PM at normal incidence (solid black line) and the maximum angle is given by the rotational discrete symmetry of the structure (red dashed line).

These peaks (maxima in SCS function) correspond to localized states, particular field patterns confined in the photonic molecule [16]. These molecular states result from the combination of individual Mie resonances in the various cylinders at a local scale and the rotational symmetry imposed by the geometry of the PM. The same electric field intensity distributions have been previously found in structures which contain decagonal and octagonal clusters of cylinders such as quasicrystals and Penrose's tilling [22,23,24]. What we observe here is the great difference in the excitation of these patterns depending on the radiation incidence angle. Peak B corresponds to a structure formed by field maxima alternating positive and negative polarization with respect the x-direction in each of the cylinders. This field distribution is excited when the radiation wavefront hits a single cylinder when impinging in the structure (0° in our reference frame). However, if the radiation finds two cylinders at the same time (at the other angular position) it is impossible to drive them with alternate polarities, making impossible the excitation of the mode. On the contrary, peak A corresponds to a field distribution where field maxima are shared by two cylinders if the octagon (decagon) is rotated 22.5° (18°), what allows very similar patterns of excitation when the radiation hits in a single cylinder at 0° as can be seen in Figures 4 and 5.

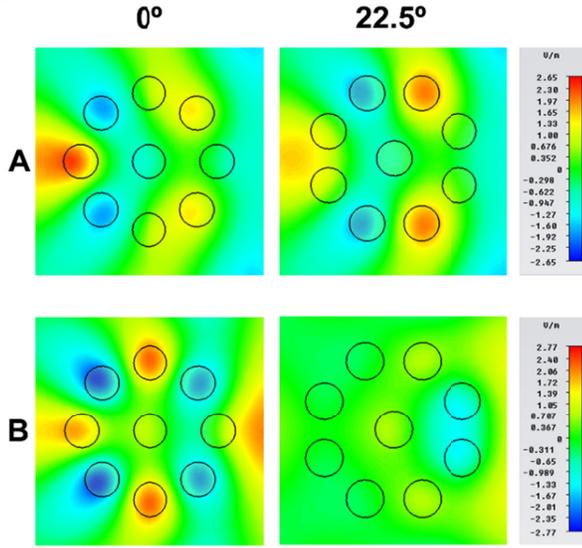

Fig. 4. Distribution of the electric-field in yz plane for peaks A and B of the octagonal PM. The red and blue colors indicate the electric field intensity and polarity, green color indicates null value for the electric field.

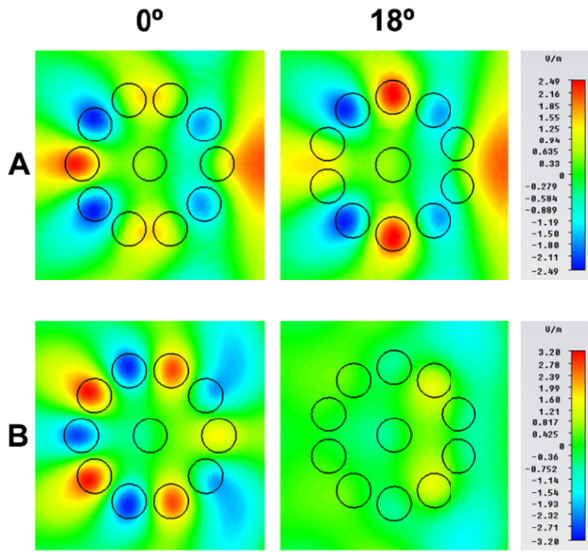

Fig. 5 Distribution of the electric-field in yz plane for peaks A and B of the decagonal PM. The red and blue colors indicate the electric field intensity and polarity, green color indicates null value for the electric field.

Finally, we measured the transmission of physical prototypes of the structures under study. The same geometrical configuration of the measurement (including the antennas) was also modeled in the simulator to obtain the corresponding transmission values [21]. It is important to note the different calculation procedure in the case of the SCS (where plane wave excites an infinite structure) and in the case of transmission measurements (where a finite sample is illuminated by finite antennas). The results of these measurements and calculations are presented in Figure 6. The Figure shows three-dimensional representations where transmission spectra (transmittance versus normalized frequency) are piled at a third axis for the different angles of the incident radiation. The transmission value is also depicted in a color scale to make easier the following of variations. As can be seen, the agreement between calculation and measurements is quite good. Besides the agreement between experimental and calculated, transmission data are in accordance with SCS data previously analyzed, not only in the profile of transmission spectra but also in the position of peaks A and B. These peaks behave as expected from the SCS, while A remains almost unchanged, B decreases significantly its value as angle varies.

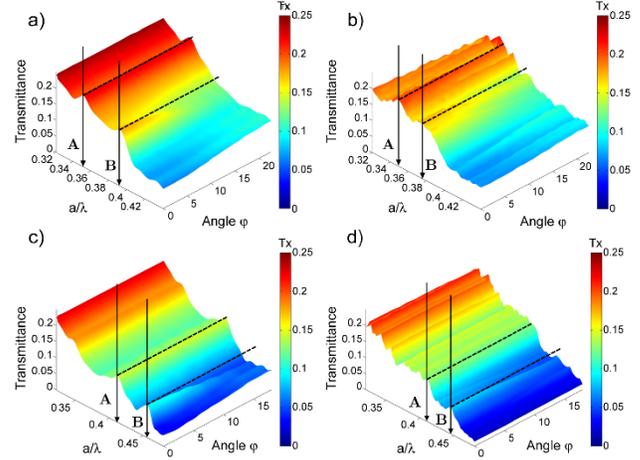

Fig. 6. Color map plots of calculated (left) and measured (right) transmission spectra for a decagonal PM as a function of normalized frequency (a/λ) and angle (φ). Upper figures (a, b) are for the octagonal ring and lower figures (c, d) for the decagonal ring. Arrows and black dashed lines show the frequency position and transmission variation of peak A and B with respect to the angle.

The existence of two states showing transmission and scattering properties so different in terms of the rotation angle φ, is an interesting feature for angle sensors design. A differential sensor can be made based on the state which is strongly dependent on the angle (B) and another state invariant under angle rotations, where the value of the angle would be determined by the difference of the transmission intensity value between the two states [15]. In general, differential operation in sensors provides multiple advantages such as systematic error cancellation, cross-sensitivities, noise reduction, etc. As a proof of concept of this idea, we have drawn a calibration curve for the octagonal structure based on data presented in Figure 6. We define ΔT as differential transmission value between frequency values of the resonances shown by the structures, ΔT=|$T_B$-$T_A$|, where $T_B$ corresponds to transmission at the frequency of resonance dependent of the azimuth angle and $T_A$ corresponds to transmission at the frequency of resonance independent of the azimuth angle. Next, we compute the values of ΔT using the calculated and measured transmission spectra of the samples. Figure 7 shows the result for the octagonal structure. These values of ΔT are plotted, normalized to ΔT at φ=0° (ΔT$_0$), in Figure 7. The agreement between the measurement and calculated data is good, as already observed in Figure 7. The calibration curve obtained present a S-shape typical in many instrumentation systems but can be linearly approximated without much error.

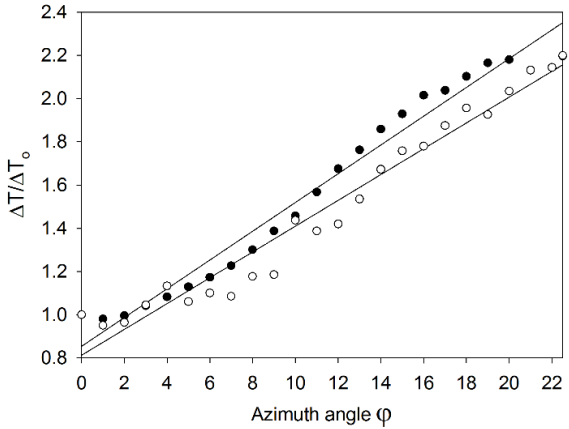

Fig. 7 Evolution of normalized differential transmission parameter $\Delta T/\Delta T_0$ in the octagonal PM as a function of the azimuth angle. Black circles correspond to calculations and white circles to measurements.

The results here presented are more general than the case studied in detail. The presented structures have been tested in the microwave range, but the scaling properties of Maxwell equations allow a direct extrapolation to other electromagnetic regimes. It is also important to use materials with the same dielectric permittivity in the desired regime. The permittivity value along this work was $\varepsilon=4.5$ of the glass rods used in the experiments, however, the same states are also found for higher values of the dielectric permittivity. For example results of $\varepsilon=13$, are shown in fig 8, where much higher Q are present.

Besides, as it can be seen in Figure 8, the states that undergo a significant dependence with respect to the angle are also present when $\varepsilon=13$, and show the same electrical field distribution as the states previously analyzed with glass rods of permittivity equal to 4.5. It is important to note that the electric field will be strongly localized inside the cylinders with a dielectric permittivity higher, and could be employed to increase the sensitivity of the structure analyzed. Therefore, the selection of the material is not a strong limitation the approach here presented to the use of photonic molecules as differential angle sensors.

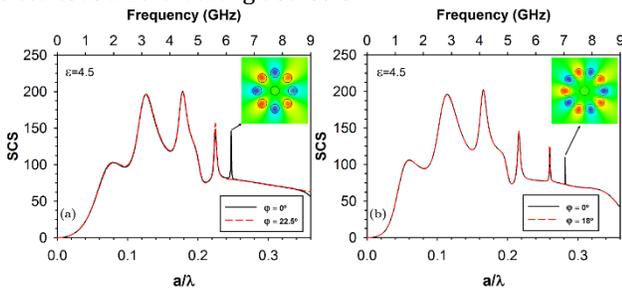

Fig. 8. Scattering Cross Section (SCS) as a function of the normalized frequency for the (a) octagonal and (b) decagonal PM with dielectric permittivity of the rods $\varepsilon=13$. The inset shows the electric field distribution of the states highly dependent on the incidence angle.

In conclusion, we have studied isolated photonic molecules made with dielectric cylinders. We found that the same resonant states shown in bulk photonic quasicrystals are present in isolated octagons or decagons. Furthermore, we have found a particular state in each type of these photonic molecules whose excitation varies strongly with the angle of incident radiation. This state shows electric field intensity maxima inside each cylinder of the ring, which corresponds to individual Mie resonances with alternating polarities. This behavior is present regardless of the dielectric permittivity of the cylinders and in the two regular structures studied (decagonal and octagonal). The agreement between measurements and simulations is good which makes these structures excellent candidates to design differential angle sensors.

**Funding.** This work has been supported by Ministerio de Economía y Competitividad of Spain via Project TEC2014-51902-C2-2-R.